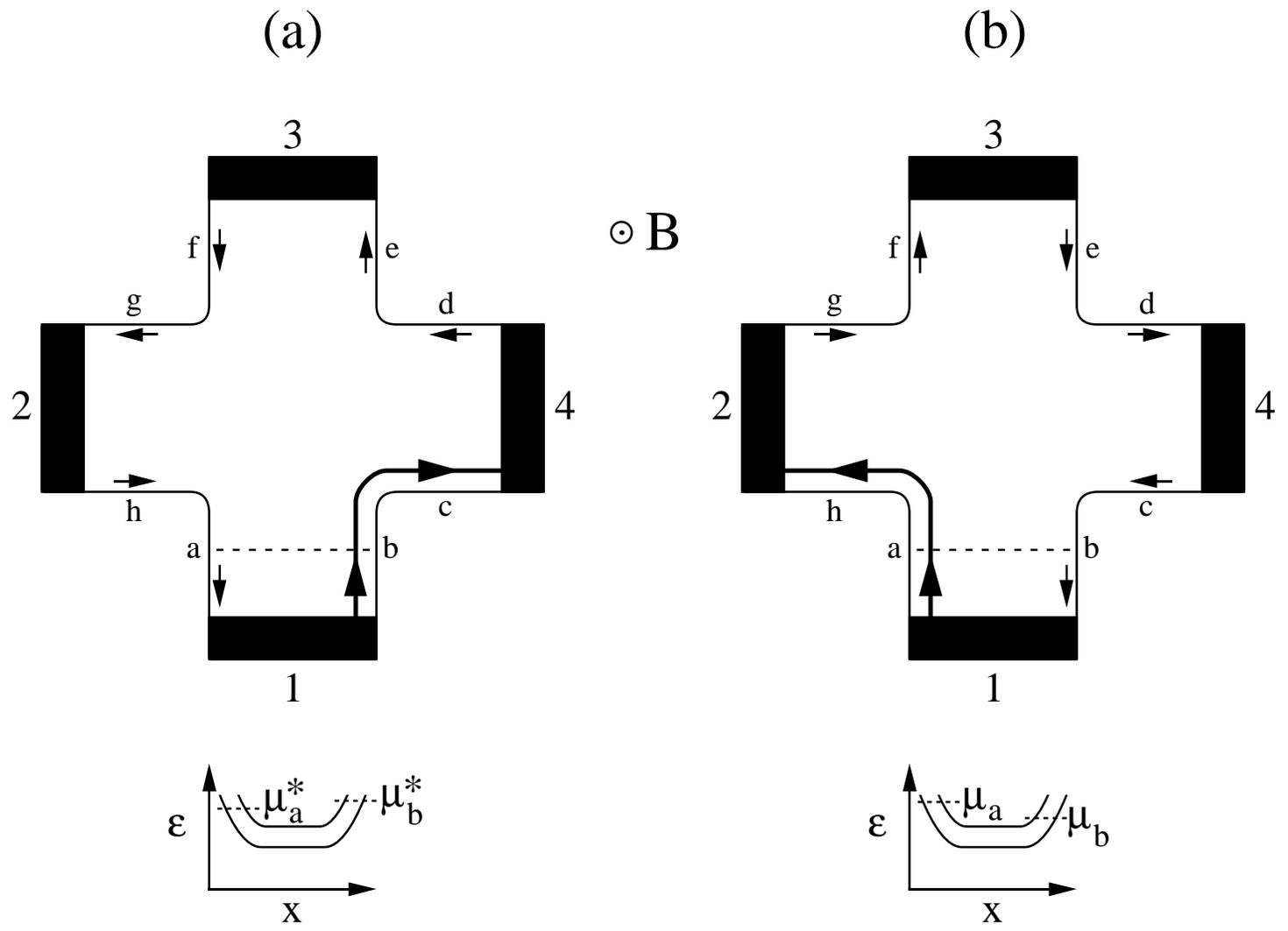

Figure 1

# D.C. Transport Measurements and the Direction of Propagation of Composite Fermion Edge States

George Kirczenow

*Department of Physics, Simon Fraser University,
Burnaby, British Columbia,
Canada, V5A 1S6*

Simple and quite general considerations are used to show that the results of d.c. transport experiments on macroscopic Hall bars are inconsistent with Hartree models in which a majority of the branches of single-particle composite fermion edge states at the Fermi level propagate in the direction opposite to that in which non-interacting electrons travel along the edge.

Some years ago Jain[1] pointed out that the fractional quantum Hall effect can be understood as the integer quantum Hall effect of composite fermions. Since then an important issue has been whether it is possible to construct a theory of the fractional quantum Hall effect based on single-particle composite fermion edge states, analogous to the very successful theories[2-5] of the integer quantum Hall effect that are based on the properties of single-particle electron edge states.

Recently a number of different models of composite fermion edge states have been proposed. In one of these models,[6] which has successfully explained the results of numerous transport experiments in the fractional quantum Hall regime,[6] all of the composite fermion single-particle edge states at a given edge propagate in one direction, the direction in which non-interacting electron edge states propagate. This direction will henceforth be referred to as the "standard direction." In the other models,[7,8] which are based on numerical Hartree calculations, single-particle composite fermion edge states can propagate in either the standard direction or the opposite direction which will be referred to as the "reverse direction." For example, for systems with a Landau level filling fraction $\nu=2/3$ in the bulk, the Hartree models predict that at edges with hard wall potentials *all* of the single-particle composite fermion edge states at the Fermi level propagate in the *reverse* direction. For soft wall $\nu=2/3$ edges the Hartree models predict that different branches of single-particle composite fermion edge states can propagate in different directions, but that at the Fermi level there are more branches that propagate in the reverse direction than in the standard direction.[8]

It has been suggested by Chklovskii and Halperin[9] that because composite fermion edge electrochemical potentials can differ from electron edge electrochemical potentials (as was pointed out earlier in Ref. 6), this surprising result of the Hartree theories does not conflict with pulsed experiments[10] that have observed the direction of propagation of edge magneto-plasmons. These are *collective* excited states whose propagation direction is not necessarily the same as that of the single-particle composite fermion edge states. However, whether models in which the majority of branches of *single-particle* composite fermion edge states at the Fermi level propagate in the reverse direction are consistent with the known results of *d.c.* Hall experiments (in which edge mag-neto-plasmons are not excited), was not considered. This is addressed in the present note.

Let us suppose, to start with, that all of the single-particle composite fermion edge states at the Fermi level propagate in the reverse direction as in the Hartree models of a $\nu=2/3$ hard edge,[7,8] and compare this scenario with the accepted edge state picture of the *integer* quantum Hall effect[2-5] in which all of the electron edge states travel in the standard direction. These two situations are shown schematically in Fig.1(a) and Fig.1(b), respectively. In Fig.1 the black areas represent contacts. In Fig.1(b) the arrows indicate the (standard) direction of propagation of single-particle electron edge states. In Fig.1(a) the arrows show the (reverse) direction of propagation predicted by the Hartree models for the single-particle composite fermion edge states.

Suppose that initially each of the two systems shown in Fig.1 is at equilibrium so that the electron electrochemical potentials of all of its four contacts are equal. Now let us increase the electron electrochemical potential of contact 1 in each device slightly (and adiabatically) while keeping the electron electrochemical potentials of the other three contacts fixed. It is well established experimentally (and is a consequence of the second law of thermodynamics) that this results in a net flux of electrons out of contact 1 and into the device. For the case in Fig.1(a) this also implies that there must be a net flux of composite fermions into the device from contact 1 because the transformation from electrons to composite fermions is a gauge transformation and therefore preserves both charges and electric currents.

Because of the direction of propagation of the electron states at the edges, in the case of Fig.1(b) the net flux of electrons from contact 1 into the device must be associated with a rise of the electron electrochemical potential $\mu_a$ along edge $a$ of the device as is illustrated in the lower part of Fig.1(b). This results in an unbalanced flux of electrons flowing into the device along edge $a$, which is indicated by the heavy directed line in the figure. This unbalanced electron flux continues on, following edge $h$ into contact 2. On the other hand, because of the assumed direction of propagation of composite fermion edge states in Fig.1(a), the situation there is different: The net flux of composite fermions coming from contact 1 must be achieved by a rise of the composite fermion electrochemical potential $\mu_b^*$ along edge $b$. Thus the unbalanced composite fermion flux in Fig.1(a) must flow along edge $b$ and continue along edge $c$ to contact 4. [Note that a net flux of composite fermions out of contact 1 would also occur if the composite fermion electrochemical potential along edge $a$ in Fig.1(a) were lowered instead of that along edge $b$ being raised. However, this way of realizing a net flux out of contact 1 in response to the change in the electrochemical potential of contact 1 would be unphysical since the composite fermions along edge $a$ originate at contact 2, and therefore their electrochemical potential should be determined by the conditions at contact 2 and not at contact 1].

Now consider a Hall resistance measurement in which contact 1 is the electron source (as in the above discussion), contact 3 is the electron drain, and contacts 2 and 4 are the Hall voltage probes which draw no net current. Experimentally (or once again as a consequence of the second law of thermodynamics) the net electron flux out of a contact increases (algebraically) when the electron electrochemical potential of the contact is raised. Therefore in order to satisfy the condition that contacts 2 and 4 act as voltage probes and draw no net current in the above thought experiment, it is necessary to raise the electron electrochemical potential of contact 2 in Fig.1(b) and of contact 4 in Fig.1(a). That is, the Hall voltage generated in Fig.1(a) has the *opposite* sign to that generated in Fig.1(b).

This means that the Hall voltage that is generated in a composite fermion system in which all of the single-particle composite fermion edge states propagate in the reverse direction has the opposite sign to that generated in systems in the integer quantum Hall regime. The same result is true for macroscopic composite fermion systems in which *a majority* of the branches of single-particle composite fermion edge states at the Fermi level propagate in the reverse direction. This is because an edge carrying $N$ standard and $M$ reverse channels with $M > N$ behaves like an edge with $M - N$ reverse channels and no standard channels if the edge is of macroscopic length so that the channels that it carries can equilibrate, a result proved in the theory of quantum railroads.[11]

Thus the prediction of the Hartree theories that for a ν=2/3 edge (whether soft or hard) a majority of the branches of single-particle composite fermion edge states at the Fermi energy propagate in the reverse direction, implies that the Hall voltage for a macroscopic ν=2/3 device should have the opposite sign to that observed in the integer quantum Hall regime. It is very well established experimentally however that the sign of the Hall voltage is *the same* at ν=2/3 (and at the other fractional Landau level fillings) as it is in the integer quantum Hall regime.[12]

It is therefore evident that the predictions that the Hartree theories make about the direction of propagation of the single-particle composite fermion edge states for ν=2/3 conflict with the results of d.c. transport experiments. This disagreement is also present for other Landau level filling fractions for which the composite fermion effective magnetic field is antiparallel to the true magnetic field in the bulk of the sample. On the other hand there is no such disagreement between experiments and the predictions of the model of composite fermion edge states proposed in Ref. 6.

I wish to thank B. L. Johnson, C. J. B. Ford and D. B. Chklovskii for interesting discussions. This work was supported by NSERC of Canada.

**Figure Caption:**

**Fig.1** Schematic drawing of two Hall bars. The black areas are contacts. The arrows show the direction of propagation for single-electron edge states in the integer quantum Hall regime in Fig.1(b), and the majority direction of propagation for composite fermion single-particle edge states predicted by the Hartree models at $\nu=2/3$ in Fig.1(a). The heavy directed curves indicate the path of the unbalanced flux of electrons (Fig.1(b)) and composite fermions (Fig.1(a)) when the electron electrochemical potential of contact 1 is increased. The plots at the bottom show schematically the electron (Fig.1(b)) and composite fermion Hartree (Fig.1(a)) Landau level structure along the dashed lines *ab* in the respective Hall bars depicted above and the corresponding electron and composite fermion edge electrochemical potentials $\mu$ and $\mu^*$, respectively.